\documentclass{aip-cp}

\usepackage[numbers]{natbib}
\usepackage{rotating}
\usepackage{graphicx}
\usepackage{epstopdf}
\usepackage[usenames,dvipsnames]{color}
\usepackage[percent]{overpic}
\newcommand*{\myfont}{\fontfamily{phv}\selectfont}

\begin{document}

\title{Detailed VHE Studies of the Pulsar Wind Nebula HESS J1825-137}

\author[aff1]{A.~M.~W. Mitchell\corref{cor1}}
\author[aff2]{C. Mariaud}
\author[aff1]{P. Eger}
\author[aff3]{S. Funk}
\author[aff1]{J. Hahn}
\author[aff1]{J. Hinton}
\author[aff1]{R.~D. Parsons}
\author[aff1]{V. Marandon}
\author{\textit{for the H.E.S.S. Collaboration}}

\affil[aff1]{Max-Planck-Institute f\"{u}r Kernphysik, Saupfercheckweg 1, 69117 Heidelberg, Germany}
\affil[aff2]{Laboratoire Leprince-Ringuet, \'{E}cole Polytechnique, CNRS/IN2P3, F-91128 Palaiseau, France}
\affil[aff3]{Friedrich-Alexander-Universit\"{a}t Erlangen-Nurnberg, Erlangen 
 Centre for Astroparticle Physics, Erwin-Rommel-Str. 1, D 91058 Erlangen, Germany}
\corresp[cor1]{Corresponding author: Alison.Mitchell@mpi-hd.mpg.de}

\maketitle

\begin{abstract}
The pulsar wind nebula (PWN) HESS~J1825-137, known to exhibit strong energy dependent morphology, was discovered by HESS in 2005 \cite{Funk06,Aharonian05pwn1825}. Powered by the pulsar PSR~B1823-13, the TeV gamma-ray emitting nebula is significantly offset from the pulsar. The asymmetric shape and 21~kyr characteristic age of the pulsar suggest that HESS~J1825-137 is in an evolved state, having possibly already undergone reverse shock interactions from the progenitor supernova. Given its large angular extent, despite its 4~kpc distance, it may have the largest intrinsic size of any TeV PWN so far detected. A rich dataset is currently available with H.E.S.S., including H.E.S.S. II data with a low energy threshold, enabling detailed studies of the source properties and environment. We present new views of the changing nature of the PWN with energy, including maps of the region and spectral studies. 
\end{abstract}

\section{INTRODUCTION}
\label{sec:intro}
HESS~J1825-137 (RA: 18 26 13.175, Dec: -13 34 46.8) is a highly extended pulsar wind nebula (PWN), discovered as part of the first H.E.S.S. Galactic plane survey \cite{Aharonian05pwn1825}. The nebula is powered by the Vela-like pulsar PSR~B1823-13; which has a spin-down power of roughly $3\times 10^{36} \mathrm{ergs}^{-1}$, a characteristic age 21 kyr, and is situated at a distance of approximately 4 kpc. HESS~J1825-137 exhibits strong energy dependent morphology, as demonstrated in a previous dedicated publication \cite{Funk06}. The observed size of the nebula decreases with increasing energy, becoming more compact around the position of the pulsar. 

The H.E.S.S. experiment is an array of IACTs situated in the Khomas Highlands of Namibia. Ten years on from the previous dedicated HESS~J1825-137 publication, the currently available H.E.S.S. dataset is much enhanced, and together with improved analysis procedures, current studies are significantly more sensitive than the previously published data to weaker emission regions.

\section{H.E.S.S. DATA}
\label{sec:data}
The H.E.S.S. data is divided into two datasets, A and B, according to the configuration of telescopes used in the analysis. Dataset A comprises the four original H.E.S.S. telescopes, CT1-4, including about 300 hours of data taken between 2004 and 2015. Dataset B consists only of data including CT5, the newest telescope of the H.E.S.S. array, which became operational in 2012. This dataset comprises about 40 hours with CT1-5, taken between 2012 and 2015. All data were analysed stereoscopically, i.e. more than one telescope triggered on each event. For this analysis, a sensitive likelihood-based analysis procedure was used (ImPACT \cite{Parsons14}).
The energy threshold of H.E.S.S. II on this dataset is around 120 GeV, much lower than the previous threshold of 270 GeV \citep{Funk06}.

\subsection{ENERGY RESOLVED ANALYSIS}
\label{sec:emaps}
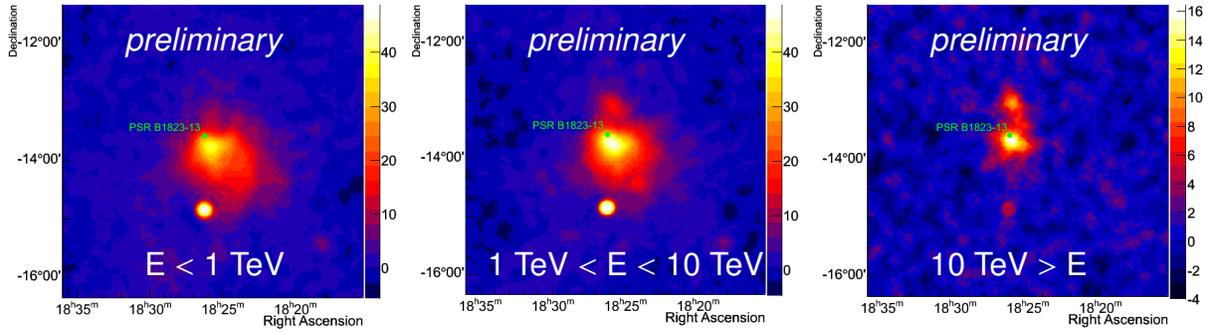
\begin{figure*}
\centering
\begin{overpic}[width=0.32\columnwidth]{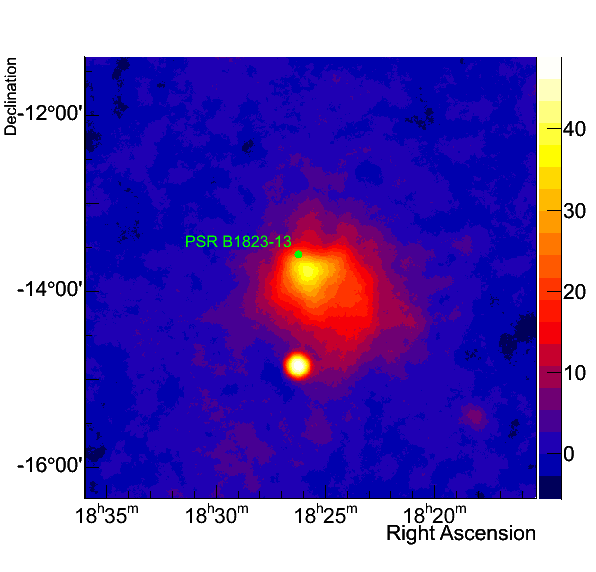}
\put(30,75){{\myfont\large{\textcolor{white}{\textit{preliminary}}}}}
\put(35,18){{\myfont\large{\textcolor{white}{E $< $ 1 TeV}}}}
\end{overpic}
\begin{overpic}[width=0.32\columnwidth]{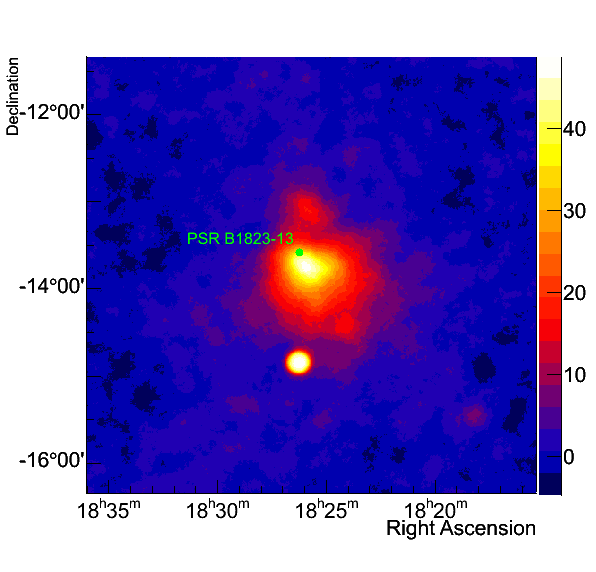}
\put(30,75){{\myfont\large{\textcolor{white}{\textit{preliminary}}}}}
\put(20,18){{\myfont\large{\textcolor{white}{1 TeV $<$ E $<$ 10 TeV}}}}
\end{overpic}
\begin{overpic}[width=0.32\columnwidth]{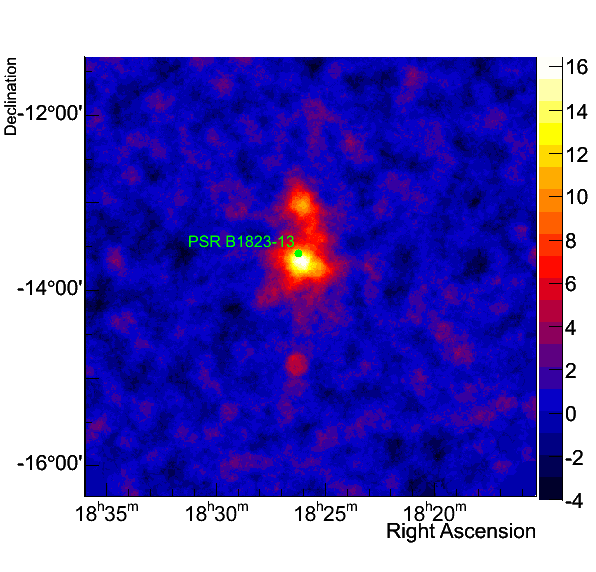}
\put(30,75){{\myfont\large{\textcolor{white}{\textit{preliminary}}}}}
\put(30,18){{\myfont\large{\textcolor{white}{10 TeV $>$ E}}}}
\end{overpic}
\caption{Significance maps of the HESS~J1825-137 region in three different energy bands, produced using dataset A. The size of the sources is clearly much reduced at high energies. Other sources within the field of view include the binary LS~5039 and the hard spectrum source HESS~J1826-130.}
\label{fig:Ebands}
\end{figure*}

The longer dataset A (using CT1-4) was analysed in three energy bands, from which the nebula size can be clearly seen to reduce with increasing energy, as shown in Fig. \ref{fig:Ebands}. This is clear evidence of the emission being attributable to the pulsar PSR~B1823-13, and provides some indication of cooling of the electron population over time as the particles are transported away from the pulsar. To the North of HESS~J1825-137, an additional source is seen which becomes more significant towards higher energies; HESS~J1826-130 \cite{gamma1826}.
 Towards the South, the point-like source LS~5039, a gamma-ray binary system, also prominently features \cite{gamma5039}. 
 The peak of the nebula emission is also seen to be offset from the pulsar position.

\begin{figure*}
\begin{center}
 \begin{overpic}[width=0.45\textwidth]{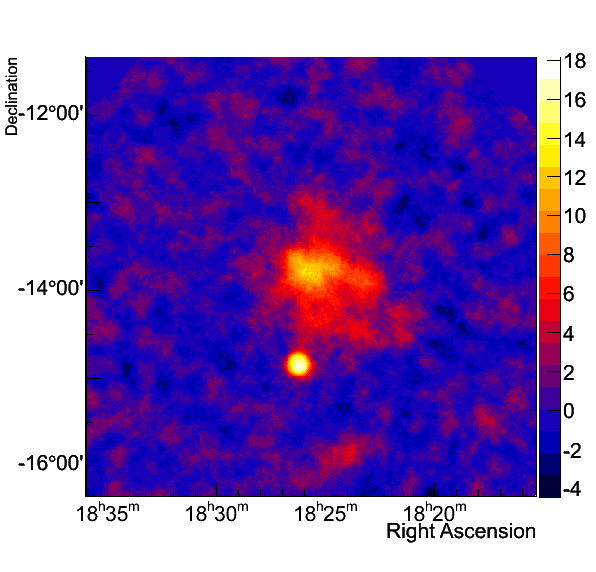}
 \put(30,75){{\myfont\large{\textcolor{white}{\textit{preliminary}}}}}
 \end{overpic}
    \caption{H.E.S.S. II era map, using post 2012 data with CT1-5 in a stereoscopic analysis (dataset B).}
    \label{fig:Bsigmap}
\end{center}
\end{figure*}

A significance map of the same region made using dataset B is shown in figure \ref{fig:Bsigmap}. This map was produced using a stereoscopic analysis of data from CT1-5; the full H.E.S.S. II array. Due to the reduced exposure time, the area of significant emission is clearly much reduced with respect to Fig. \ref{fig:Ebands}. By contrast, a significance map made using the full dataset A and saturated at 20 sigma demonstrates how with increased exposure, the analysis is more sensitive to areas of weaker emission. The total apparent area of the nebula is visibly increased with respect to that of the shorter lifetime dataset B shown in Fig. \ref{fig:Bsigmap}.

\subsection{NEBULA SPECTRUM}
\label{sec:spectrum}
\begin{figure*}
\begin{center}
\begin{overpic}[width=0.45\textwidth]{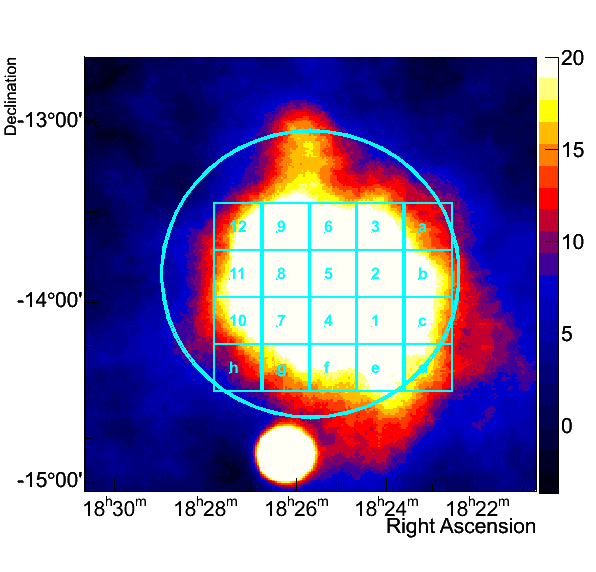}
\put(20,80){{\myfont\large{\textcolor{white}{\textit{preliminary}}}}}
 \end{overpic}
  \caption{Saturated map showing the regions used in the spectral analyses. The circle of $0.8^\circ$ radius was used for the nebula spectrum. Twenty spatially resolved boxes were used in Fig. \ref{fig:boxes}.}
  \label{fig:satmap}
\end{center}
\end{figure*}

Overlaid on Fig. \ref{fig:satmap} are the various regions from which a spectral analysis was performed. Given the extensive angular size of the nebula, it is possible to conduct spatially resolved spectral analyses, from different regions of the nebula independently.

\begin{figure*}
\begin{center}
\begin{overpic}[trim=0 2cm 0 0,clip=true,width=0.6\textwidth]{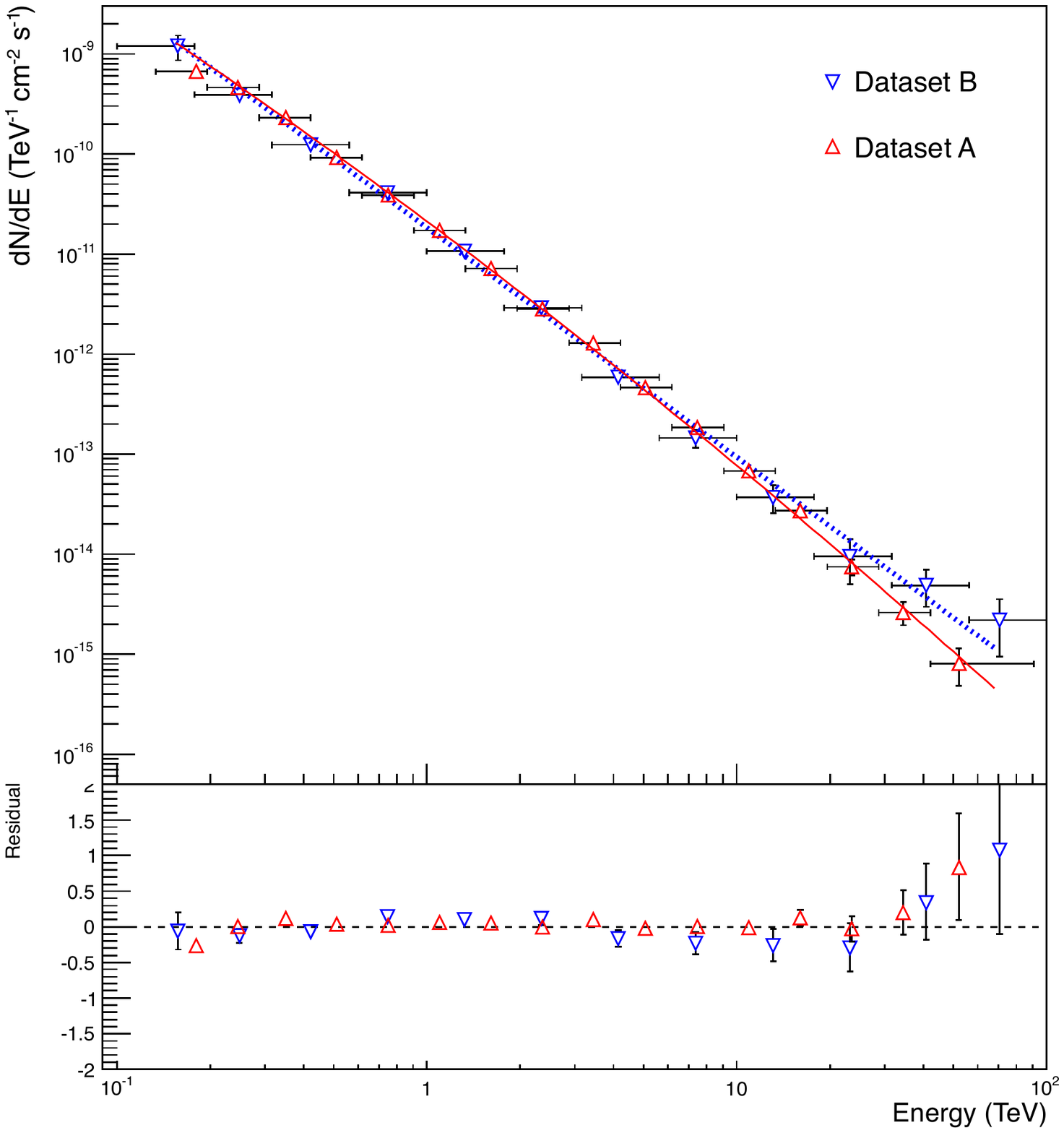}
\put(25,40){{\myfont\large{\textcolor{black}{\textit{preliminary}}}}}
 \end{overpic}
\caption{A spectrum taken from the circular region shown in Fig. \ref{fig:satmap}, encompassing the majority of the nebula. Parameters are in good agreement with the previously reported spectrum \cite{Funk06}. See Table \ref{tab:spectralfits}. A best fit log parabola is shown for dataset A (with a $\chi^2$/ndf of 32/13)  and a power law fit for dataset B.}
\label{fig:spectrum}
\end{center}
\end{figure*}

A spectrum was extracted from the same circular region (shown in Fig. \ref{fig:satmap}) as was used in \citep{Funk06}, encompassing the majority of the nebula. This is shown in Fig. \ref{fig:spectrum}; in dataset A, curved models are preferred over a power law fit, whereas curved models are not significant for dataset B. Whilst dataset B has a lower energy threshold than dataset A, the high energy emission is more significant in the longer dataset A. 

\subsection{SPATIALLY RESOLVED SPECTRAL ANALYSIS}
\label{sec:boxes}

\begin{table*}
\begin{center}
\caption{Fit parameters for various fits to the nebula spectrum. Curved models are preferred for dataset A, fitted in the energy range [0.133, 91] TeV. A power law fit is quoted for the shorter dataset B, in the energy range [0.12, 91] TeV.}
\label{tab:spectralfits}
\tabcolsep7pt\begin{tabular}{cccc}
\hline
  $\mathrm{d}N/\mathrm{d}E$ Fit Model & $I_0$ (1TeV) / $10^{-11}\mathrm{cm}^{-2}\mathrm{s}^{-1}$  & $\Gamma$  & Fit Parameters  \\
\hline
$I_0 E^{-\Gamma}$ & $1.99\pm 0.02$ & $2.32\pm 0.01$ & -   \\
$I_0 E^{-\Gamma}\exp(-E/E_c)$ & $2.07\pm 0.02$ & $2.21\pm 0.01$ & $E_c = 24\pm 4$TeV  \\
$I_0 E^{-\Gamma + \beta\log E}$ & $2.12\pm 0.02$ & $2.31\pm 0.01$ & $\beta= 0.057 \pm 0.007$  \\
$I_0 E^{-\Gamma}$ (dataset B) & $1.83\pm 0.06$ & $2.29\pm 0.03$ & - \\
\hline
\end{tabular}
\end{center}
\end{table*}

A grid of $0.26^\circ\times 0.26^\circ$ boxes laid over the nebula was used to conduct a spatially resolved spectral analysis. As the box sizes are larger than the H.E.S.S. point-spread-function, the spectra obtained from each region are independent.

\begin{figure*}
\begin{center}
 \begin{overpic}[trim=0 2cm 0 2cm,clip=true,width=0.47\textwidth]{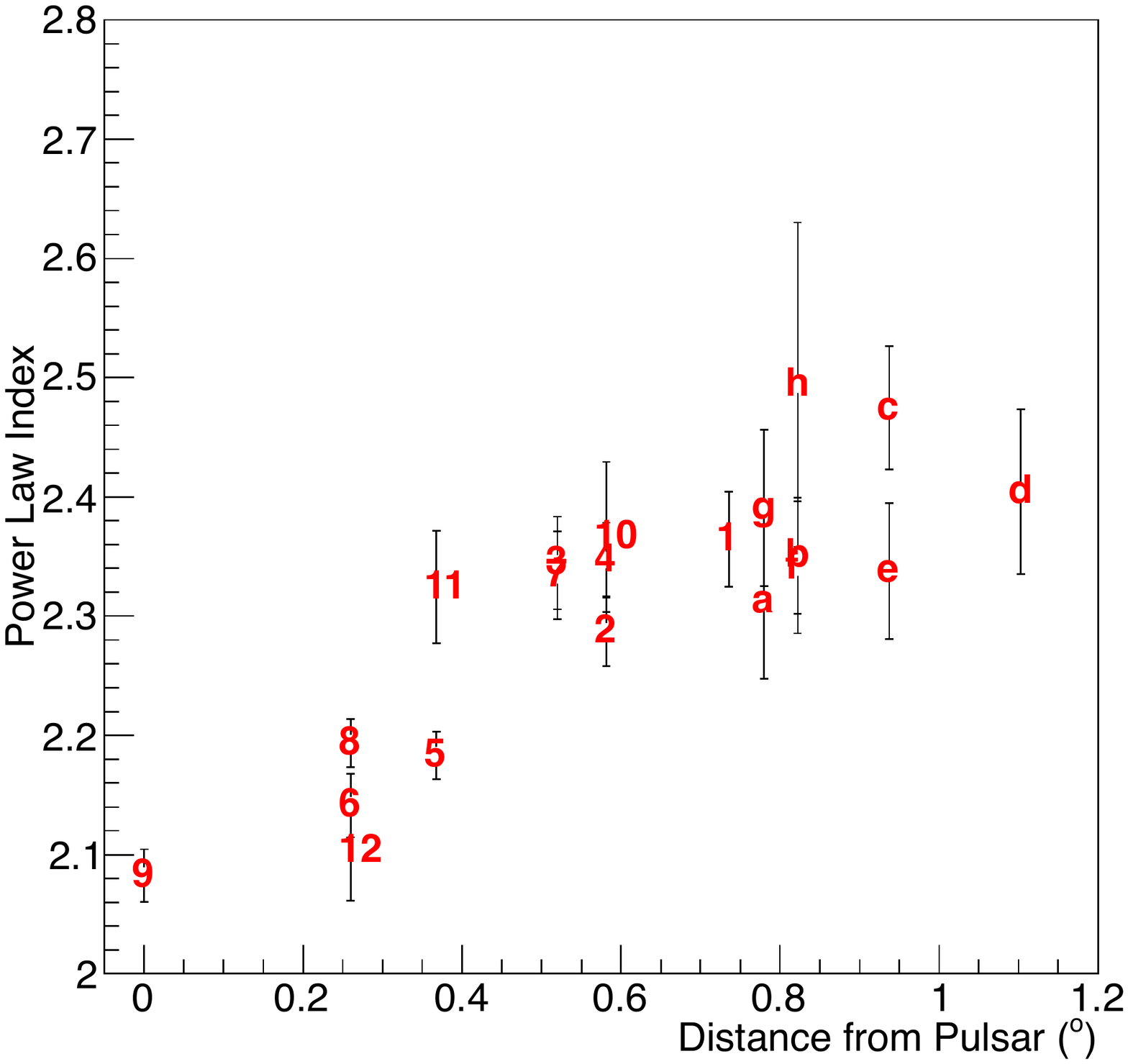}
 \put(20,70){{\myfont\large{\textcolor{black}{\textit{preliminary}}}}}
 \end{overpic}
 \begin{overpic}[trim=0 2cm 0 2cm,clip=true,width=0.47\textwidth]{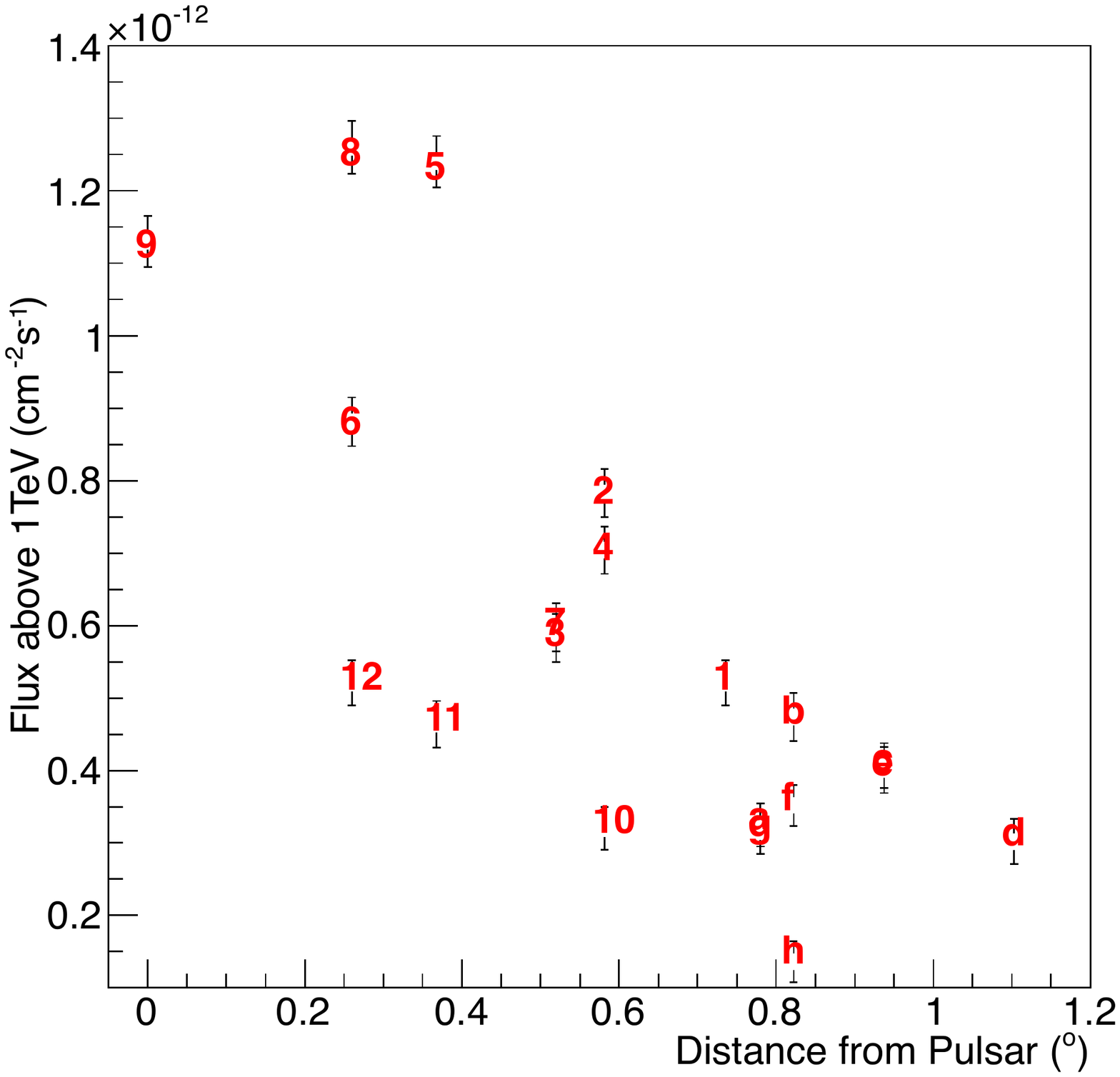}
\put(50,70){{\myfont\large{\textcolor{black}{\textit{preliminary}}}}}
 \end{overpic}
  \caption{A spatially resolved spectral analysis. Clearly, the flux decreases (left) and spectral index softens (right) with increasing distance from the pulsar. There are 8 new regions beyond $0.7^\circ$.}
\label{fig:boxes}
\end{center}
\end{figure*}

Spectral regions 1-12, shown in Fig. \ref{fig:satmap}, were previously analysed in \cite{Funk06}. In the present analysis, eight newly significant regions beyond $0.7^{\circ}$ from the pulsar (a-h) were additionally included, together with a re-analysis of regions 1-12, using dataset A. These additional regions have become sufficiently significant thanks to the increased exposure, adding testimony to the increased size of the HESS~J1825-137 nebula with respect to that previously reported. The spectra extracted from each of these 20 regions were separately fitted with a power law; the variation in power law index and in flux above 1 TeV as a function of distance from the pulsar are shown in Fig. \ref{fig:boxes}. The spectral indices are subject to an additional systematic error of 0.1; whilst only statistical errors are shown in Fig. \ref{fig:boxes}. The spectral index clearly softens with distance from the pulsar, the two variables have a correlation coefficient of $r = 0.8 \pm 0.1$. Conversely, the flux decreases with distance from the pulsar, with a correlation coefficient of $r = -0.75 \pm 0.02$. 

From both the spatially resolved spectral analysis presented here and the energy resolved maps in Fig. \ref{fig:Ebands}, it can be seen that the lower energy emission extends further away from the pulsar than the high energy emission. This is generally interpreted as due to the cooling of Inverse Compton producing electrons as they propagate away from the pulsar. As the electrons are transported away from the pulsar, energy is lost through cooling processes including adiabatic expansion of the nebula and synchrotron cooling losses. Gradually, the energy losses lead to a shift in the electron spectrum towards lower energies, reflected in the consequent $\gamma$-ray emission. 

\section{CONCLUSIONS}
\label{sec:conclude}
HESS~J1825-137 extends further than has been previously seen at TeV energies. The additional area of extended emission has been revealed due to a better sensitivity to large areas of weaker, lower energy emission with the current analysis. 
This improvement is due to a much longer dataset and more powerful analysis tools being currently available. Strong energy dependent morphology is exhibited across the nebula. The rich dataset now available, both from the extended livetime of dataset A and the extended energy range of dataset B, enables detailed studies of the nebula properties. In agreement with previous findings, the spectral index of the emission increases with increasing distance from the pulsar, due to the electrons cooling over time, causing the index to become softer. Additionally, the high energy flux decreases with distance from the pulsar, due also to this gradual change in the energy distribution of the electron population, as they cool and are transported through the nebula.




\section{ACKNOWLEDGMENTS}
The support of the Namibian authorities and of the University of Namibia in facilitating the construction and operation of H.E.S.S. is gratefully acknowledged, as is the support by the German Ministry for Education and Research (BMBF), the Max Planck Society, the German Research Foundation (DFG), the French Ministry for Research, the CNRS-IN2P3 and the Astroparticle Interdisciplinary Programme of the CNRS, the U.K. Science and Technology Facilities Council (STFC), the IPNP of the Charles University, the Czech Science Foundation, the Polish Ministry of Science and Higher Education, the South African Department of Science and Technology and National Research Foundation, the University of Namibia, the Innsbruck University, the Austrian Science Fund (FWF), and the Austrian Federal Ministry for Science, Research and Economy, and by the University of Adelaide and the Australian Research Council. We appreciate the excellent work of the technical support staff in Berlin, Durham, Hamburg, Heidelberg, Palaiseau, Paris, Saclay, and in Namibia in the construction and operation of the equipment. This work benefited from services provided by the H.E.S.S. Virtual Organisation, supported by the national resource providers of the EGI Federation.

\nocite{*}
\bibliographystyle{aipnum-cp}%
\bibliography{J1825gamma16}%

\end{document}